\documentclass[prd,amsmath,11pt]{revtex4}
\usepackage[dvips]{graphicx}
\begin{document}
\title{Approximate solutions for the single soliton \\ in a Skyrmion-type model with a dilaton scalar field}
\author{J. A. Ponciano$^{\dag\; \flat}$ and C. A. Garc\'{\i}a Canal$^{\ddag}$}
\affiliation{$^{\dag}$Departament de F\'{\i}sica Te$\grave{o}$rica, Universitat de  Val$\grave{e}$ncia-CSIC, E-46100, Burjassot (Valencia), Spain \\
$^{\flat}$Departamento de F\'{\i}sica, Universidad de San Carlos, cd. universitaria Z.12, Guatemala \\
$^{\ddag}$Laboratorio de F\'{\i}sica Te\'orica, Departamento de F\'{\i}sica, Universidad Nacional de La Plata, IFLP-CONICET, C.C.67 (1900) La Plata, Argentina}

\vspace{1cm}
\begin{abstract}
\hfill

We consider the analytical properties of the single soliton solution in a Skyrmion-type Lagrangian that incorporates the scaling properties of quantum chromodynamics (QCD) through the coupling of the chiral field to a scalar field interpreted as a bound state of gluons. The model was proposed in previous works to describe the Goldstone pions in a dense medium, being also useful for studying the properties of nuclear matter and the in-medium properties of mesons and nucleons.
Guided by an asymptotic analysis of the Euler-Lagrange equations, we propose approximate analytical representations for the single soliton solution  in terms of rational approximants exponentially localized. Following the Pad\'e method, we construct a sequence of approximants  from the exact power series solutions near the origin. We find that the convergence of the approximate representations to the numerical solutions is considerably improved by taking the expansion coefficients as free parameters and then minimizing the mass of the Skyrmion using our ans\"atze for the fields. We also perform an analysis of convergence by computation of physical quantities showing that the proposed analytical representations are very useful for further phenomenological calculations.

\end{abstract}

\maketitle

\section{Introduction}
Through the formalism of effective field theory a significant progress has been reached in our attempt to undertstand the physics involving mesons and baryons in nuclear matter. The first effective descriptions appeared as generalizations of the linear $\sigma$ model so that the physical constraint dictated by the broken-realized chiral symmetry of QCD was naturally included. More recently, the interest was focused on models incorporating the broken scale invariance of strong interactions and therefore suitable for describing physics beyond the pion sector \cite{Serot,Tang}.
 Among these models, an interesting approach consists in introducing a scalar glueball with an appropriate potential into a nonlinear realization of the $\sigma$ model, thus accounting for both broken-scale and chiral symmetries \cite{Karliner}.
From this background, the authors in Ref. \cite{Lee} proposed a Skyrmion-type Lagrangian with spontaneously broken chiral and scale symmetries. This model, which is valid as an approximation to QCD in the large $N_c$-limit, is useful for studying the properties of pions in dense medium and can also describe multibaryon systems, including dense hadronic matter. In this scenario, baryons arise as solitonic solutions of the equations of motion that, because  of their nonlinearity, cannot be solved in an analytical closed form.

Naturally, it is desirable that the advance in the theoretical domain be accompanied by the solution of the mathematical problems introduced. Our contribution in this paper is
to give analytical approximate representations for the single soliton that is used to define the parameters of the theory at zero density. This types of representation are preferred to numerical solutions because with them one can trace exact analytical information about the soliton solution. Besides, explicit representations are very convenient for phenomenological calculations in the analysis of dense Skyrmion matter. Our procedure will consists of extracting the structure of the exact solution from an asymptotic analysis of the equations of motion for the scalar and chiral fields, and then we look for solutions in terms of rational approximants, Pad\'e Approximants, that have proved to be particularly successful in giving a faithful representation of the topological configuration of the original Skyrme Model \cite{Ponciano}. Here, the task is more involved since it requires the resolution of two coupled nonlinear differential equations for the new scalar field and the chiral field already present in the Skyrme Model. Furthermore, the nonzero mass for pions and the additional scalar field imply exponential asymptotic behaviors for both solutions, and this feature must be incorporated in the approximants. We will see that the analytical approach to the single Skyrmion with the scalar field problem can be implemented in a simple way and that
the suggested representations reproduce the properties of the exact solutions for small and large values of the radius $r$.

Our paper is organized as follows. In the next section we briefly describe the model that is composed by the original Skyrme Model implemented with the trace anomaly of QCD, signal of the broken-realized scale invariance. In Sec.III we present the properties of the single Skyrmion and the scalar field and then, in Sec.IV we construct the rational representations by using the exact asymptotic solutions for both the chiral and the scalar field and we present an analysis of convergence of the proposed approximants. Finally, we give the conclusions of our work in Sec.V.


\section{The Model}
The model constitutes an extension to the original Skyrme Model that emerged as an attempt to describe baryons interacting by means of meson exchanges \cite{Witten}. In a more recent formulation introduced by Ellis and Lanik \cite{Ellis} and developed in further works for the description of nuclear matter and finite nuclei \cite{Brown}, the task was aimed at incorporating the underlying scale invariance of QCD under the scale transformation
\begin{eqnarray}
x\rightarrow \lambda^{-1}x, \quad \lambda \ge 0.
\end{eqnarray}

 This symmetry is broken at the quantum level by dimensional transmutation, and its breaking is manifested through a nonvanishing trace of the energy momentum tensor $\theta_{\mu\nu}$, referred to as the QCD trace anomaly which in the chiral limit is given by
\begin{equation}
\theta_{\mu}^{\mu}=\frac{\beta(g)}{2g}\mathrm{Tr}G_{\mu\nu}G^{\mu\nu},
\end{equation}
where $\beta (g)=-(g^3/48\pi^2)(11N_c-2N_f)$ is the one-loop beta function of $QCD$.

The description starts with the Skyrme Lagrangian for massive pion fields, which reads
\begin{equation}
\mathcal{L}=\frac{f_{\pi}^2}{4}\mathrm{Tr}(\partial_{\mu}U^{\dagger}\partial^{\mu}U)+\frac{1}{32e^2}\mathrm{Tr}[U^{\dagger}\partial_{\mu}U,U^{\dagger}\partial_{\nu}U]^2+\frac{f_{\pi}^2m_{\pi}^2}{4}\mathrm{Tr}(U+U^{\dagger}-2),
\end{equation}
where the Goldstone pion fields $\pi^a$  play the role of a chiral phase angle present in the chiral field $U=\exp(i\mbox{\boldmath $\tau$}\cdot\mbox{\boldmath $\pi$}/f_{\pi})$ belonging to $SU(2)$, and $f_{\pi}$, $m_{\pi}$ and $e$ are the pion decay constant, the pion mass and the Skyrme parameter, respectively.

The broken-scale invariance of QCD is implemented by the introduction of a scalar field $\chi(r)$ with a scale dimension 1. This scalar degree of freedom is interpreted as a bound state of gluons, the glueball field, that operates on two-pion exchange physics in meson-exchange phenomenology.

With an appropriate scaling breaking-potential for the scalar field $\chi(r)$, the modified Lagrangian that respects the scaling properties of QCD reads
\begin{eqnarray}\label{model}
\mathcal{L}&=&\frac{f_{\pi}^2}{4}\left(\frac{\chi}{f_{\chi}}\right)^2\mathrm{Tr}(\partial_{\mu}U^{\dagger}\partial^{\mu}U)+\frac{1}{32e^2}\mathrm{Tr}([U^{\dagger}\partial_{\mu}U,U^{\dagger}\partial_{\nu}U])^2+\frac{f_{\pi}^2m_{\pi}^2}{4}\left(\frac{\chi}{f_{\chi}}\right)^3\mathrm{Tr}(U+U^{\dagger}-2) \nonumber \\
&&+\frac{1}{2}\partial_{\mu}\chi\partial^{\mu}\chi-\frac{1}{4}\frac{m^2_{\chi}}{f^2_{\chi}}\left[\chi^4\left(\ln(\chi/f_{\chi})-\frac{1}{4}\right)+\frac{1}{4}\right].
\end{eqnarray}
Here, $f_{\chi}$ is the vacuum expectation value of the scalar field $\chi(r)$. The mass of $\chi(r)$ is given by $m^2_{\chi}=d^2V(\chi)/d\chi^2$, where $V(\chi)$ denotes the scale-breaking potential. No experimental values are available for those parameters, and conjectural values depend rather on specific considerations of the model, as it is explained in Ref. \cite{Vento}.  The inability to fix them comes from the lack of knowledge about the mixing between the two components of the trace anomaly corresponding, respectively, to the chiral (``soft'') and scale (``hard'') symmetry-breaking contributions. Mixing depends on the baryonic sector to be described. For instance, in matter it is assumed that the glueball contribution decouples gradually with increasing density, reaching a critical point at which  only the ``soft'' component is relevant.  In such a case, the parameters can be fixed constraining the model by the bulk properties of finite nuclei and yielding the values $f_{\chi}=240$ MeV and $m_{\chi}=550$ MeV. In section IV, we present the approximants to the solitonic solutions corresponding to the different sets of parameters reported elsewhere.
Concerning the pion parameters, the accepted experimental values are $f_{\pi}=93$ MeV and $m_{\pi}=140$ MeV.

\section{Properties of the single Skyrmion with the Scalar Field}
In this section we analize the static solutions with nontrivial topology of the Euler-Lagrange equations of the model. We first present the exact asymptotic solutions, and then we propose suitable analytical continuations to these representations.

Let us first derive the Euler-Lagrange equations from the Lagrangian (\ref{model}). Adopting the usual hedgehog ansatz \cite{Balachandran} for the chiral field $U=\exp(i\mbox{\boldmath $\tau$}\cdot\mbox{\boldmath $\pi$}/f_{\pi})$ and assuming also spherical symmetry for the additional scalar field \cite{Vento}, that is
\begin{equation}
U(r)=\exp(i\mbox{\boldmath $\tau$ }\cdot \mathbf{r}F(r)),\qquad \chi(r)=f_{\chi}C(r),
\end{equation}
we can write the mass of the single soliton in the Skyrme model as
\begin{eqnarray}\label{mass}
E[F,C]&=&4\pi \int_0^{\infty}r^2dr \left\{\frac{f^2_{\pi}}{2}C^2\left[\left(\frac{dF}{dr}\right)^2+2\frac{\sin^2F}{r^2}\right]+\frac{1}{2e^2}\frac{\sin^2F}{r^2}\left[\frac{\sin^2F}{r^2}+2\left(\frac{dF}{dr}\right)^2\right]\right.+ \nonumber \\
&+&\left. f_{\pi}^2m^2_{\pi}C^3(1-\cos F)+\frac{f_\chi^2}{2}\left[\left(\frac{dC}{dr}\right)^2+\frac{m_{\chi}^2}{2}\left(C^4(\ln C-1/4)+\frac{1}{4}\right)\right]\right\}.
\end{eqnarray}

Variations of $E[F,C]$ with respect to the profile function $F(r)$ and the scalar field $C(r)$ lead to the classical equations of motion.

In terms of the dimensionless variable $y$, defined as $y=ef_{\pi}r$, the coupled equations for the fields $F$ and $C$ read
\begin{eqnarray}\label{motionF}
(y^2C^2&+&2\sin^2F)\frac{d^2F}{dy^2}+2\left(yC^2+y^2C\frac{dC}{dy}\right)\frac{dF}{dy} + 2\sin F\cos F \left(\frac{dF}{dy}\right)^2 +\nonumber \\&-&2C^2\sin F \cos F -2 \frac{\sin^3F\cos F}{y^2}-\left(\frac{m_{\pi}}{ef_{\pi}}\right)^2y^2C^3\sin F=0,
\end{eqnarray}
and,
\begin{eqnarray}\label{motionC}
\frac{d^2C}{dy^2}+\frac{2}{y}\frac{dC}{dy}-\left(\frac{f_{\pi}}{f_{\chi}}\right)^2C\left(\left(\frac{dF}{dy}\right)^2+2\frac{\sin^2F}{y^2}\right)
-3 \left(\frac{m_{\pi}}{ef_{\chi}}\right)^2C^2(1-\cos F)
-\left(\frac{m_{\chi}}{ef_{\pi}}\right)^2C^3 \ln C =0. \nonumber \\
\end{eqnarray}

Finiteness of energy requires that $U(r)$ tends to an arbitrary constant element of $SU(2)$ at spatial infinity. Choosing $U(r)\rightarrow 1$ as $r\rightarrow \infty$ implies, for the chiral angle $F(r)$, the boundary condition
\begin{equation}
F(\infty)=0.
\end{equation}
This condition compactifies the space of constant time surfaces, $R^3$, to the three-sphere $S^3$; so the field $U(r)$ defines a map from the compactified configuration space $S^3$ to the identity in the target space, $SU(2)$, which is isomorphic to $S^3$. The space of the distinct equivalence classes by homotopy corresponds to the group
\begin{equation}
\Pi_3[SU(2)]=\Pi_3(S^3)=Z,
\end{equation}
which means that the model has an infinite number of topological sectors, each one characterized by an integer-valued winding number. Z is defined as the topological charge $q=\int dx^3 B_0$, which is related to the baryonic number current defined as
\begin{equation}\label{Bcurrent}
B^{\mu}=\frac{1}{24\pi^2}\epsilon^{\mu\nu\rho\sigma}Tr[(U^{\dagger}\partial_{\nu}U)(U^{\dagger}\partial_{\rho}U)(U^{\dagger}\partial_{\sigma}U)].
\end{equation}

As for the original Skyrme model, the equations of motion for $F(y)$ and $\chi(y)$ admit power-series solutions, of the form
\begin{eqnarray}\label{powerseries}
F(y)&\sim& \pi +F_1 y + F_3 y^3 + O(y^5), \nonumber\\
C(y)&\sim& C_0 +C_2 y^2 + O(y^4),
\end{eqnarray}
which are valid for small values of $y$. The zero-order coefficient from the power-series representation of $F(y)$ is choosed equal to $\pi$ in order to constrain the solution to the $B=1$ sector. The coefficients $F_1=-\alpha$ and $C_0$ cannot be determined by the perturbative analysis.

By substitution of the power-series solutions (\ref{powerseries}) into Eqs.(\ref{motionF}) and (\ref{motionC}), the following recurrence relations among the coefficients $F_1$, $F_3$, $C_0$ and $C_2$ are found:

\begin{eqnarray}\label{coefficients}
C_2&=&\frac{1}{2}\left(\frac{f_{\pi}}{f_{\chi}}\right)^2 C_0F_1^2+\frac{m_{\pi}^2}{e^2f_{\chi}^2}C_0^2-\frac{1}{6}\frac{m^2_{\chi}}{e^2f_{\pi}^2}C_0^3\ln C_0,\\
F_3&=&-\frac{1}{30}\frac{F_1}{e^2f_{\pi}^2(C_0^2+2F_1)^2}(2F_1^4e^2f_{\pi}^2+3m_{\pi}^2C_0^3+12C_0C_2e^2f_{\pi}^2+4C_0^2F_1^2e^2f^2_{\pi}). \nonumber
\end{eqnarray}

On the other hand, from the asymptotic behavior of the equations of motion it can be shown that the functions $F$ and $C$ reach their vacuum values at infinity following the functional forms:
\begin{equation}\label{asymptotic}
F(y)\sim \frac{\exp(-m_{\pi}y/ef_{\pi})}{y}, \qquad C(y)\sim 1-\frac{\exp(-m_{\chi}y/ef_{\pi})}{y}.
\end{equation}

Our task in the next section will is to find approximate analytical representations for the functions $F(r)$ and $C(r)$, valid within the whole domain of definition of the radius $r$.

\section{Approximate representations}

Previous works \cite{Linde,Ananias} that addressed solving the equation of motion of the original Skyrme Model have shown the utility, for practical calculations, of implementing an approach that incorporates both asymptotic behaviors of the chiral angle $F(r)$, near the origin and at large $r$, in one single representation. The rational approximants have proved to be suitable for this purpose. Moreover, it has been shown in Ref.\cite{Ponciano} that starting from the well-known Pad\'e approximant approach, a simple and systematic method to solve the Skyrme problem can be developed.
In the following discussion we extend the analytical approach for the single Skyrmion with the scalar field, showing that the exponential behavior (\ref{asymptotic}) can be easily incorporated into rational representations, providing satisfactory analytical continuations of the power-series representations (\ref{powerseries}). We will begin by computing a sequence of Pad\'e approximants using suitable ans\"atze for the functions $F(r)$ and $C(r)$ suggested from the previous study of the stucture of the exact solution.
We continue the discussion by showing that the convergence to the exact solution is drastically improved when using a variational method to minimize $E[F,C]$, supplemented by our ans\"atze for the fields. Then the attention is focused on analysing the convergence of the proposed solutions by use of some typical physical observables of Skyrmion-type models.

For the sake of completness we briefly remind the reader that Pad\'e approximants are rational functions used to give an analytic continuation of a power-series representation of a given function \cite{Baker}. The Pad\'e approximant $P_{[M,N]}(y)$ of order $[M,N]$ to the series $S(y)=\sum_n a_ny^n$ is defined as the ratio of two polynomials
\begin{equation}
P_{[M,N]}(x)=\frac{\sum_{k=0}^{M} a_ky^k}{1+\sum_{k=1}^N b_k y^k},
\end{equation}
where the free $N+M+1$ parameters, $a_k, b_k$, are fixed so that the first $M+N+1$ coefficients in the Taylor expansion of $P_{[M,N]}$ coincide with the series $S(y)$ up to order $M+N$.

In the present problem, modified approximants for the chiral angle $F(r)$ may be built so as to match the exponential asymptotic behavior given in relation (\ref{asymptotic}).

Working with the dimensionless variable $y$ defined previously, we may cast the approximate solutions for $F$ in the form
\begin{equation}\label{ansatz}
F(y)=\frac{\sum_{k=0}^ma_ky^k}{1+\sum_{k=1}^{m}b_ky^k+b_{m+1}y^{m+1}\exp(m_{\pi}y/ef_{\pi})},
\end{equation}
where the parameters $a_k$ and $b_k$ are to be fixed so as to reproduce the behavior of the exact solution near the origin.

Note that the functional form that is proposed has the exact asymptotic behavior at infinity, that is, $F(r)\sim \exp(-m_{\pi}r)/r$.

The lowest order solution that can be built is
\begin{equation}\label{F[0,1]}
F(y)=\frac{\pi}{1+(\alpha /\pi)y\exp(m_{\pi}y/ef_{\pi})}.
\end{equation}
The above approximant contains the asymptotic behavior, $\exp(-m_{\pi}y/ef_{\pi})/y$, and agrees with the power-series solution near the origin up to order $O(y)$.
It is possible to construct another approximant of the same order [0,1], but with an additional parameter $b_2$,
\begin{equation}
F(y)=\frac{\pi}{1+\frac{(\alpha-b_2\pi)}{\pi}y+b_2y\exp(m_{\pi}y/ef_{\pi})},
\end{equation}
$b_2$ should be chosen so as to reproduce the exact asymptotic decay for large $r$.

The higher-order approximants contain the information of additional terms of the exact expansion at the origin. Obviously, they improve the degree of approximation to the exact solution. For instance, keeping the ansatz (\ref{ansatz}) and fixing the coefficients by means of the Pad\'e approximant method using equations (\ref{powerseries}-\ref{coefficients}), we find for the $[1,2]$ approximant the following expression
\begin{eqnarray}\label{AP[1,2]}
F[1,2]=\frac{\pi(1+a_1y)}{1+b_1y+b_2y^2\exp(m_{\pi}y/f_{\pi}e)}
\end{eqnarray}
with
\begin{equation}
a_1=-\frac{\alpha}{\pi}+\frac{\pi F_3}{\alpha^2-\pi\alpha m_{\pi}/ef_{\pi}},\qquad b_1=\frac{\pi F_3}{\alpha^2-\pi\alpha m_{\pi}/ef_{\pi}}, \qquad
b_2=\frac{F_3}{\alpha-\pi m_{\pi}/e f_{\pi}},
\end{equation}
and for the $[2,3]$ order approximant
\begin{equation}\label{AP[2,3]}
F[2,3]=\frac{\pi(1+a_1y+a_2y^2)}{1+b_1y+b_2y^2+b_3y^3\exp(m_{\pi}y/ef_{\pi})}
\end{equation}
with 
\begin{eqnarray}
a_1=-\frac{\alpha}{\pi}+2(F_3^2+\alpha F_5)(\pi \xi-\alpha)/\Delta,
\quad a_2=\left[ (\alpha-\pi \xi)\left(2\alpha(F_3^2+\alpha F_5)-\pi F_3^2\right)-2\pi^2F_3F_5\right]/\Delta, \nonumber
\end{eqnarray}
\begin{eqnarray}
b_1=2(F_3^2+\alpha F_5)(\pi \xi-\alpha)/\Delta,\quad b_2=\xi F_3(\pi \xi-2\alpha)-2\pi F_5/\Delta ,\quad b_3=-2F_3(F_3^2+\alpha F_5)/\Delta
\end{eqnarray}
where $\xi=m_{\pi}/ef_{\pi}$ and $\Delta=F_3[2\pi F_3+\alpha \xi(\pi\xi-2\alpha)]$. \\

By construction, from the Taylor expansion of the Pad\'e approximants (\ref{AP[1,2]}) and (\ref{AP[2,3]}) we recover the exact behavior near the origin [relations (\ref{powerseries})] up to orders $O(y^3)$ and $O(y^5)$, respectively. Note also that, as the proposed approximants satisfy the right boundary conditions, they yield to topological solutions of baryonic number equal to $q=1$. That is, recalling Eq.(\ref{Bcurrent}),
\begin{eqnarray}
q&=&\int dx^3B_0 \\ \nonumber
&=&-\frac{1}{2\pi^2}\int_0^{\infty}\frac{dF}{dy}\frac{\sin ^2 F}{y^2}4\pi y^2dy =-\frac{1}{\pi}\left[F(y)-\frac{\sin 2F(y)}{2}\right]_0^{\infty}=1.
\end{eqnarray}

The sequence of Pad\'e approximants to the chiral angle $F(y)$ given by equations (\ref{F[0,1]}-\ref{AP[2,3]}) is shown in Fig.1 together with the exact numerical solution. The approximate representations show convergence to the exact numerical solutions; unfortunately,  we found that higher-order calculations in the Pad\'e procedure lead to singular approximants being no longer relevant to represent $F(r)$.

In order to eliminate spurious representations with singularities in the real $y$ axis one may proceed heuristically by adding restrictions to the numerator coefficients of the approximants. For instance, by simply setting $a_k=1$ for all $k$ we will produce well-behaved and positive-definite approximants, as shown in Fig.2. This procedure, which relies on simplicity, can be particularly useful in problems for which one seeks for an analytical  description of Skyrmion-Skyrmion interactions such as in dibaryon models or other multi-Skyrmion models \cite{Schroers}. However the prescription of setting $a_k=1$ is too arbitrary and does not really select the solutions that minimize the energy functional $E[F,C]$.

One can actually confirm this fact by studying the sequence of restricted rational approximants $[m,m+1]$ for the chiral angle.
The sequence of restricted Pad\'e approximants is displayed in Fig.2 where we can see a fast convergence to the $[4,5]$ order representation. As we should expect, the convergence to the numerical solution is very fast at the boundaries, however the representations underestimate in $1\%$ to $2\%$ in the $1\lesssim y\lesssim 2$ region. Higher order approximations slightly improve the agreement with the numerical solution but at the price of getting less attractive expression for using in phenomenological calculations from the model.
Quality tests of the approximation are done further in this work by calculating physical quantities that frequently enter in the evaluation of the static properties of the nucleons.

The low convergence to the numerical solution is a consequence of the fixing condition $a_k=1$  which appears to be too restrictive.
We will see further in this section that the convergence to the exact numerical results is drastically improved and the arbitrariness in the approximate representations eliminated by taking the expansion coefficients in the ansatz (\ref{ansatz}) as free parameters and then minimizing by numerical methods the functional of energy $E[F,C]$ given in Eq.(\ref{mass}).

To this end we first need to give a suitable ansatz for the scalar field $C(y)$. One can notice that the asymptotic behavior of $1-C(y)$ has the same functional form as the one corresponding to the profile function of the skyrmion so that analogous representations to those of $F(y)$  should provide good results for the scalar field.
Thus, we propose,
\begin{equation}\label{ansatzC}
C(y)=1-\frac{\sum_{k=0}^mc_ky^k}{1+\sum_{k=1}^{m}d_ky^k+d_{m+1}y^{m+1}\exp(m_{\chi}y/ef_{\pi})}.
\end{equation}

Of course, we should expect a change on the dependence of the rational representations parameters on the power series coefficients as, in this case, the perturbative solution is given in even powers of $y$ (see Eq.(\ref{powerseries})).

Our best representations of the exact solutions for $F(y)$ and $C(y)$ are obtained by keeping the functional forms (\ref{ansatz}), (\ref{ansatzC}) and relaxing the conditions for the expansion coefficients. These are to be found by minimizing the functional of energy following a variational method. We give the first two approximants in the sequence $[m,m+1]$ which are yet very good representations for the fields as it can be seen in Fig.3.

\begin{eqnarray}
F[1,2]&=&\frac{\pi(1-0.097y)}{(1+0.487y+0.494y^2\exp(m_{\pi}y/ef_{\pi})}, \nonumber \\
C[1,2]&=&1-\frac{(1-0.729)(1+2.003y)}{1+2.511y+0.409y^2\exp(m_{\chi}y/ef_{\pi})}, \nonumber\\
F[2,3]&=&\frac{\pi(1-0.342y+0.103y^2)}{(1+0.353y+0.171y^2+0.205y^3\exp(m_{\pi}y/ef_{\pi})}, \nonumber \\
C[2,3]&=&1-\frac{(1-0.729)(1-0.881y+0.900y^2)}{1-0.829y+1.264y^2+0.095y^3\exp(m_{\chi}y/ef_{\pi})}.
\end{eqnarray}

We now show that the analytical representations are useful for phenomenological applications by making an explicit use of them and comparing the results with numerical calculations. We have used specific values for the set of parameters $(m_{\chi},f_{\chi})$ of the effective chiral Lagrangian, namely ($720$ MeV, $240$ MeV). From Ref.\cite{Vento}, we learn that these values are relevant for the discussion of nuclear matter. The analysis for other values of $m_{\chi}$ and $f_{\chi}$, as those reported elsewhere \cite{song}, give analogous results.

We have calculated the mass of the single soliton in the Skyrme model with the scalar field given by (\ref{mass}) as well as the moment of inertia of the soliton which, as it is well known, is found after performing the collective semi-classical expansion substituting $U(r)$ by $U(r,t)=A(t)U(r)A^{\dagger}(t)$ in the Lagrangian, $A(t)$ being an $SU(2)$ matrix. For the model in question the collective transformation yields, after performing spatial integration,
\begin{equation}
L=-E[F,C] + \lambda \mathrm{Tr}[\partial_0A\partial_0A^{-1}]
\end{equation}
where $E[F,C]$ is the soliton mass and the moment of inertia $\lambda$ is given by
\begin{eqnarray}
\lambda=\frac{\pi}{3f_{\pi}e^3}\int_0^{\infty}dy\; y^2\sin^2F\left[C^2+4\left[\left(\frac{dF}{dy}\right)^2+\frac{\sin^2F}{y^2}\right]\right].
\end{eqnarray}
For the mass of the soliton and the moment of inertia we found the following numerical values $M=1387$ MeV and $\lambda=0.12/f_{\pi}$, which we take as reference values when using the analytical representations for calculations. To further check the reliability of the rational representations we have also considered the axial-vector current which enters in calculations of matrix elements within the Skyrme model.
Starting with Eq.(\ref{model}) one finds the axial-vector current to be
\begin{eqnarray}
(J_A)_{\mu}^a&=&i\frac{f_{\pi}^2}{4}\left(\frac{\chi}{f_{\chi}}\right)^2\mathrm{Tr}\left[\tau^a(\partial_{\mu}U U^{\dagger}-\partial_{\mu}U^{\dagger}U)\right]+ \\ \nonumber
& &-\frac{i}{16e^2}\left[\mathrm{Tr}\left(\left[\tau^a,\partial_{\nu}UU^{\dagger}\right]\left[\partial_{\mu}U U^{\dagger},\partial^{\nu}UU^{\dagger}\right]\right)-\mathrm{Tr}\left(\left[\tau^a,\partial_{\nu}U^{\dagger}U\right]\left[\partial_{\mu} U^{\dagger}U,\partial^{\nu}U^{\dagger}U\right]\right) \right],
\end{eqnarray}
so that the spatial integral of the axial current, expressed as a product of spatial and internal factors reads
\begin{eqnarray}
\int d^3x (J_A)_j^a=-G\mathrm{Tr}(\tau^aA\tau^jA^{-1})
\end{eqnarray}
where the factor G is explicitly given by,
\begin{eqnarray}
G=-\frac{\pi}{3e^2}\int_0^{\infty}dy\; y^2\left[C^2\left(\frac{dF}{dy}+\frac{\sin 2F }{y}\right)+4\frac{\sin 2F}{y}\left(\frac{dF}{dy}\right)^2+8\frac{\sin^2F}{y^2}\frac{dF}{dy}+ 4\frac{\sin^2F \sin 2F}{r^3}\right]. \nonumber \\
\end{eqnarray}

The values for the physical quantities obtained from the above representations are displayed in table I. We found a very good agreement between approximate and numerical solutions.

We would like to stress that the phenomenological calculations, which involve a highly nontrivial nonlinear problem, get considerably simplified through the use of the proposed analytical approximate representations. Certainly, the most interesting feature of our  representations is that they contain, in an explicit form, exact analytical properties of skyrmion solution for the Skyrme model with the scalar field problem as Eqs.(\ref{powerseries}) and (\ref{asymptotic}) can be fully recovered from them.

\begin{table}\label{table2}
\caption{Predictions from rational approximants obtained by minimizing Eq.(\ref{mass}). Masses are given in units of $f_{\pi}/e$, and $\lambda$ in units of $1/f_{\pi}$.}
\begin{tabular}{ccccc}
\hline
$\qquad$&$[1,2]$ & $[2,3]$ & $[3,4]$ & exact \\
\hline
M&$71.10$ & $70.86$ & $70.86$ & $70.87$\\
$\lambda$&$0.110$ & $0.116$ & $0.116$ & $0.116$\\
G&$0.67$ & $0.63$ & $0.64$ & $0.66$\\
\hline
\end{tabular}
\end{table}

\section{Final remarks}
The suggested approximants based on a Pad\'e-like method provide satisfactory analytical representations for the single soliton solution of an effective chiral Lagrangian, well-behaved under the scaling properties of QCD. The rational fractions proved to be well suited to incorporate the exponential behaviors of the chiral and scalar fields. Therefore they can be used with reliability to implement an analytic continuation of the exact power series solutions of the equations of motion near the origin. By this means we give an approach to a difficult problem of two coupled nonlinear differential equations where only numerical solutions are available. Besides, the analytical representations contain explicit information of the exact solution, which remains hidden in a numerical approach. Certainly, analytical representations for the fields simplify the analysis of physical situations described by the model and may be used to study the phase structure of the Skyrmion system modeling nuclear matter as well as in many other applications such as in the context of exotic baryons and monopole excitations \cite{Weigel} where the chiral soliton models are the common background.

\section*{Acknowledgements}
Many thanks to Prof. Vicente Vento for useful discussions. This work was supported by the grants GV05/264 and FPA2004-20058E. JAP acknowledges the AECI for  financial support.

\newpage

\begin{figure}[h]
\begin{center}
\includegraphics[width=10cm]{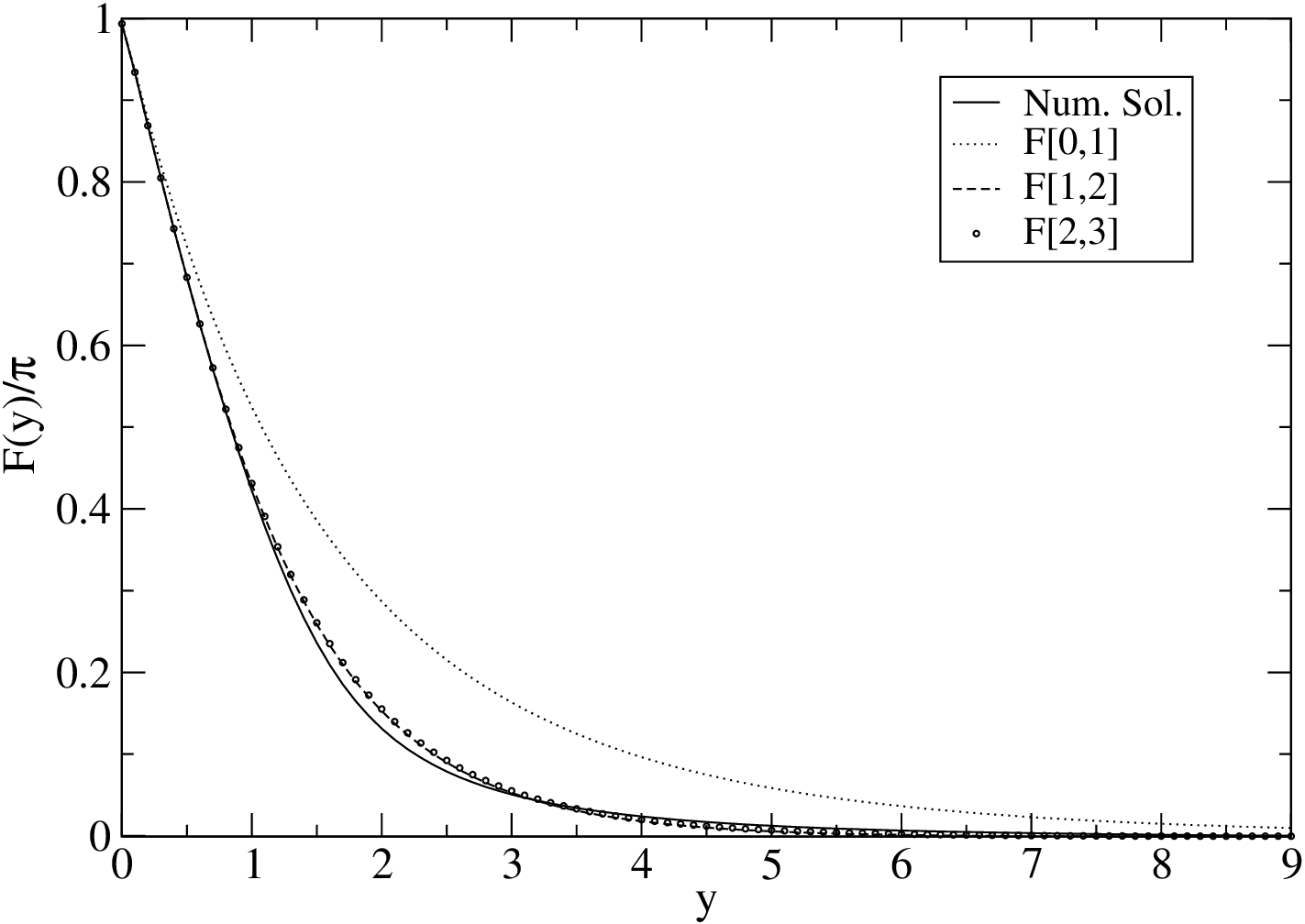}
\caption{\label{fig.1}Sequence of Pad\'e approximants to the chiral angle $F(y)/\pi$ and exact numerical solution for $m_{\chi}=720$ MeV}
\end{center}
\end{figure}

\begin{figure}[h]
\begin{center}
\includegraphics[width=10cm]{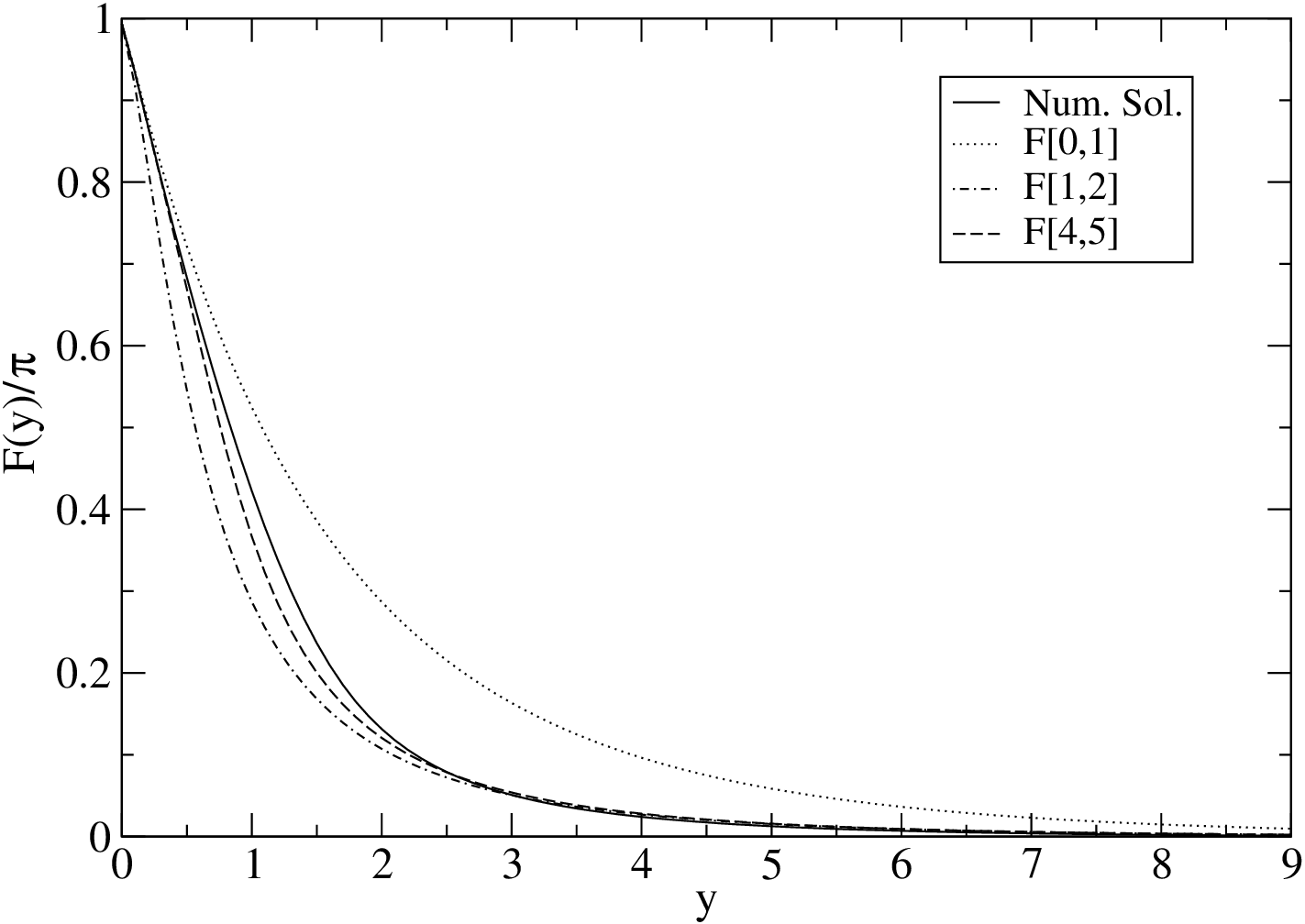}
\caption{\label{fig.2}Sequence of constrained approximants to the chiral angle $F(y)/\pi$ and exact numerical solution for $m_{\chi}=720$ MeV}
\end{center}
\end{figure}

\begin{figure}[b]
\includegraphics[width=10cm]{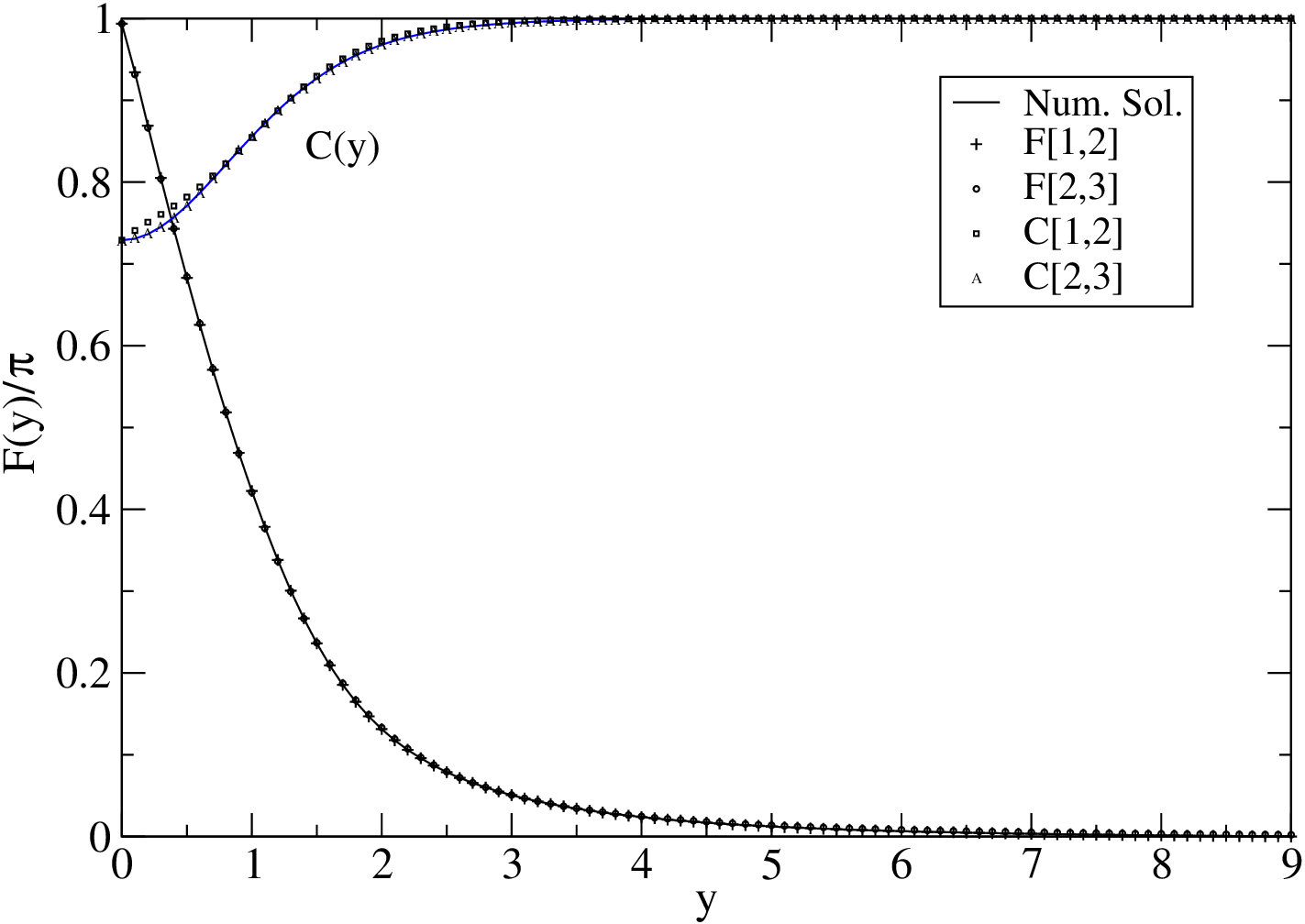}
\caption{\label{fig.3}Sequence of approximants to $F(y)/\pi$ and $C(y)$ found by minimizing the functional of energy using rational representations for the fields.}
\end{figure}
\end{document}